\documentclass[traditabstract]{aa} 
\usepackage{graphicx}
\usepackage{txfonts}

\usepackage{natbib} 
\bibpunct{(}{)}{;}{a}{}{,} 
\usepackage{url}

\newcommand{\source}{\mbox{1H 0419--577}}

\def \th {\thinspace}
\def \kms {\hbox{km s$^{-1}$}}

\def \ergsc{\hbox{erg s$^{-1}$ cm$^{-2}$}}
\def \ergsca{\hbox{erg s$^{-1}$ cm$^{-2}$ A$^{-1}$}}

\def \phsc{\hbox{photons s$^{-1}$ cm$^{-2}$}}
\def \ergs{\hbox{erg s$^{-1}$}}

\def \msun {\hbox{${\rm M_\odot}$}}

\newcommand\omegam{\hbox{{$\Omega_{\rm m}$}}}
\newcommand\omegalambda{\hbox{{$\Omega_{\Lambda}$}}}
\newcommand\kmsmpc{{\rm km s$^{-1}$ Mpc$^{-1}$}}
\newcommand\ho{\hbox{{$H_{0}$}}}
\def \nh {\hbox{ $N{\rm _H}$ }}
\def \colc {cm$^{-2}$}

\newcommand{ \lia} {Ly$ \rm{\,\sc{\alpha}}$}
\newcommand{ \fek} {Fe K$\alpha$}
\newcommand{\xmm }{{\rm XMM-\it{Newton}}}

\begin{document}

\title{Simultaneous XMM-\textit{Newton} and HST-COS observation of 1H0419-577}
\subtitle{II. Broadband spectral modeling of a variable Seyfert galaxy.}

\author{L. Di Gesu \inst{1}
      \and E. Costantini \inst{1}
      \and E. Piconcelli\inst{2}
      \and J. Ebrero\inst{1}
      \and M. Mehdipour\inst{3}
     \and J.S. Kaastra\inst{1}
}

\institute{
SRON Netherlands Institute for Space Research, Sorbonnelaan 2, 3584 CA Utrecht, The Netherlands \email{L.di.Gesu@sron.nl}
\and Osservatorio Astronomico di Roma (INAF), Via Frascati 33, I--00040, Monteporzio Catone (Roma), Italy
\and Mullard Space Science Laboratory, University College London, Holmbury St. Mary, Dorking, Surrey, RH5 6NT, UK
}

\date{Received October 30, 2013; accepted January 20, 2014}

\abstract
{
In this paper
we present the longest exposure (97 ks)
\th \xmm \th EPIC-pn spectrum ever obtained for the
Seyfert 1.5 galaxy \th \source.
With the aim of explaining
the broadband emission of this source,
we took advantage of
the simultaneous coverage in the optical/UV
that was provided in the present case
by the \th \xmm \th \th Optical Monitor and
by a HST-COS observation.
Archival FUSE flux measurements in the FUV
were also used for the present analysis.
We  successfully modeled the X-ray spectrum
together with the optical/UV fluxes data points
using a Comptonization model. 
We found that a blackbody temperature 
of $T \sim 56$ eV
accounts for the optical/UV emission
originating in the accretion disk. 
This temperature serves
as input for the Comptonized components
that model the X-ray
continuum. Both a warm ($T_{\rm wc} \sim 0.7 $ keV, $\tau_{\rm wc} \sim 7 $) 
and a hot corona ($T_{\rm hc} \sim 160 $ keV, $\tau_{\rm hc} \sim 0.5$) 
intervene to upscatter the disk photons to X-ray
wavelengths.
With the addition of
a partially covering ($C_v\sim50\%$)
cold absorber with a variable
opacity (\nh $\sim [10^{19}- 10^{22}]$ \colc),
this model can well explain also 
the historical spectral variability
of this source, with
the present dataset
presenting the lowest one
(\nh$\sim10^{19}$ \colc).
We discuss a scenario where the variable
absorber,
getting ionized in response
to the variations of the
X-ray continuum, becomes less
opaque in the highest flux states.
The lower limit
for the absorber density derived
in this scenario is typical
for the broad line region clouds.
We infer
that \th \source \th \th
may be viewed from
an intermediate inclination
angle $i \geq 54^{\circ}$,
and, on this basis,
we speculate that
the X-ray obscuration
may be associated with
the innermost dust-free 
region of the obscuring
torus. 
Finally, 
we critically  compare this scenario
with all the different models 
(e.g. disk reflection)
that have been used in the past
to explain the variability
of this source.}
\titlerunning{Simultaneous XMM-\textit{Newton} and HST-COS observation of 1H0419-577}
\authorrunning{L. Di Gesu et al}
 \maketitle


\section{Introduction}
\label{intro}
The infalling of
matter onto
a supermassive black hole
(SMBH) supplies the energy
that
active galactic nuclei
(AGN) emit in the
form of observable
radiation
over a broad energy
range.
Indeed, in the radio quiet case,
the AGN emission ranges
mostly from optical to X-ray wavelengths. 
The optical/UV emission
is thought to be direct
thermal emission from the accreting
matter. A standard
geometrically thin, optically
thick, accretion disk
\citep{sha1973,nov1973}
produces a multicolor
blackbody spectrum,
whose effective
temperature
scales
with the black hole mass
as $\propto M^{-1/4}$. 
Therefore, for a typical
AGN hosting
a supermassive black hole
(SMBH) 
of $M \sim 10^{8} \msun$,
the disk spectrum is
expected to peak
in the far UV range
(e.g., for an Eddington ratio of
$L/L_{\rm Edd} \sim 0.2 $,
 $T \sim 20 \, \rm eV$ ).\\ 
The Wien tail of the accretion
disk emission is not expected
to be strong in the soft X-rays.
However, Seyfert X-ray spectra
display a prominent 
``soft excess''
\citep{arn1985, pir1997}
that lies well above the
steep power law which
well describes the spectrum
at energies larger than $\sim$ 2.0
keV \citep{per2002, cap2006, pan2008}.
The origin of the soft X-ray emission
in AGN has been debated a lot in the
last decades.   
Comptonization of the disk photons
in a warm plasma is a 
possible mechanism
to extend 
the disk emission 
to higher energies
\citep{mag1998,don2012}.
Alternatively,
relativistically blurred 
reflection of the
primary X-ray power law in
an ionized disk
\citep{bal2001, ros2005}
is another possible
explanation.
Partial covering ionized
absorption
\citep[see][for a review]{tur2009}
is also able to
explain the
soft X-ray emission
without requiring
extremely relativistic
conditions in the vicinity
of the black hole.
Discriminate among these
models through X-ray spectral
fitting alone is difficult,
even in high signal to noise
spectra. 
In many cases, 
models with drastically different underlying
physical assumptions can provide
an acceptable fit
\citep[e.g.,][]{mid2007, cru2006}.
For instance, in
the case of the 
well known 
nearby Seyfert 1
MCG-6-30-15 
both
a reflection model
\citep{bal2003}
and an absorption
based 
model \citep{mil2008}
have been successfully used
to fit
the spectrum.
For these reasons,
the nature of the
soft excess is still an open
issue
\citep[see also][]{pic2005}.\\
Recently,
the multiwavelength monitoring
campaign (spanning over $\sim$ 100 days) 
of the bright Seyfert
galaxy Mrk 509 \citep{kaa2011}
provided a possible
discriminating evidence.
During the campaign,
the
soft excess component
varied
together with the UV continuum
emission
\citep[][hereafter M11]{meh2011}.
This finding disfavors disk reflection,
at least as the main driver of
the source variability
occurring on the few days timescale
typical of the campaign.
Indeed,
in the case of  disk reflection,
the soft excess component
should rather vary
in correlation
with the hard ($\ga 10$ keV)
X-ray flux,
because
of the broad reflection
bump at  $\sim$ 30 keV
characterizing any
reflection component.
The M11 result is not
a unique case: in a different case,
the soft excess variability
has been found to be independent
from the hard X-ray variability
also on a longer 
($\sim$ years) timescale
\citep[e.g. Mrk 590, ][]{riv2012}.\\
The correlation between
the soft X-ray and UV
variability may be
a natural consequence of
Comptonization, 
because in this framework 
the soft X-ray emission
is directly 
fed by the disk photons.
Indeed,
this interpretation
explains the simultaneous broadband
optical/UV/X-ray/gamma-ray
spectrum of Mrk 509 obtained
in the monitoring campaign
\citep[][hereafter P13]{pet2013}.\\
In the P13 broadband model,
two Comptonized
components model
respectively the
''soft-excess`` and
the X-ray emission
above $\sim$ 2.0 keV.
Indeed,
it is commonly
accepted that
the phenomenological power law
characterizing
AGN spectra above $\sim$ 2.0 keV
is produced by Comptonization
of the disk photons 
in a hot (T$\sim$ 100 keV), 
optically thin
corona \citep{haa1991}.
The nature and the origin
of the hot corona are
still largely unknown.
However, 
recent results
from X-ray timing techniques
\citep[e.g. X-ray reverberation lag,][]{wil2013}
or imaging of gravitationally lensed quasar
\citep[e.g.,][]{mor2008} indicates that
it may be a compact emitting spot, 
located a few gravitational radii
above the accretion disk
\citep[e.g.,][]{rei2013}.\\ 
%
%
\begin{figure}[t]
\centering
\includegraphics[angle=90,width=0.5\textwidth]{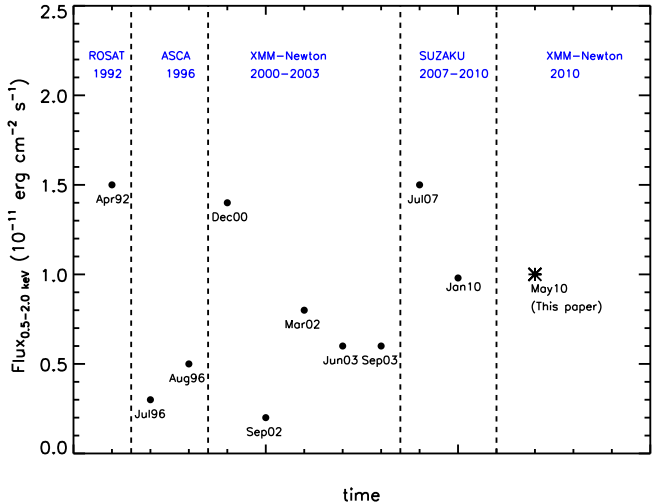}
\caption{Historical X-ray (0.5--2.0 keV) fluxes of \th \source.
Each symbol indicates a flux measurement, obtained in the labeled date.}
\label{fluxes.fig}
\end{figure}

%
%


\begin{table*}[t]
\caption{\xmm \th \th datasets log.}
\centering
\begin{tabular}{lcccccc}
\hline
\hline
Obs & Observation ID & Date & Orbit & Net exposure \tablefootmark{a} & 
$\rm F_{0.5-2.0\,keV}$ \tablefootmark{b} & $\rm F_{2.0-10.0\,keV}$\tablefootmark{b}\\
& & &  & ks & $10^{-12}\,\ergsc$ & $ 10^{-12}\,\ergsc$ \\
\hline
1\tablefootmark{c} & 0112600401 & 04/12/2000 & 181 & 6 & 14 & 16 \\
2\tablefootmark{c} & 0148000201 & 25/09/2002 & 512 & 11 & 1.8 & 8.2\\
3\tablefootmark{c} & 0148000401 & 30/03/2003 & 605 & 11 & 7.6 & 11.3  \\
4\tablefootmark{c} & 0148000501 & 25/06/2003 & 649 & 6 & 6.4 & 10.3  \\
5\tablefootmark{c} & 0148000601 & 16/09/2003 & 690 & 11  & 5.6  & 10.0  \\
\hline \hline
6\tablefootmark{d} & 0604720301/401 & 28-30/05/2010 & 1917/1918 & 97 & 10  & 14  \\
\hline
\end{tabular}
 \tablefoot{
 \tablefoottext{a}{Net exposure time after flaring filtering.}
 \tablefoottext{b}{Observed fluxes in the quoted bands}
 \tablefoottext{c}{Archival datasets.}
 \tablefoottext{d}{Datasets analyzed in this paper.}
  }
\label{archive.tab}
\end{table*}


\section{\source: a variable Seyfert galaxy }
\label{sou}

\th \source \th \th
is a radio quiet quasar
located at redshift z=0.104
and spectrally classified
as type 1.5 Seyfert \citep{ver2006}.
The estimated mass for the
SMBH harbored in its nucleus
is 
$\sim 3.8 \times 10^{8} \rm M_{\odot}$
\citep{one2005}. The source
has been targeted by all
the major X-ray observatories, and
in Fig. \ref{fluxes.fig} we 
plot the historical
fluxes in the 0.5--2.0 keV band.
As noticed for the
first time in \citet{gua1998},
\th \source \th \th undergoes
frequent transitions
between low and
high flux states.
While the bulk of the flux variation
occurs in the soft X-ray, 
in the hard (2--10 keV) 
X-ray band the 
spectral slope 
flattens out drastically
\citep [down to $\Gamma=1.0$ in the lowest state,][]{pag2012,pou2004a}.
Due to this peculiar behavior
in the X-rays, \th \source \th \th
is challenging for
any interpretation, and for this
reason it has been subject
of discussion over the past years.
\\ 
According to \citet{pag2002},
the cooling 
of the plasma temperature 
in the hot corona may
produce the observed flux/spectral
transition. Afterwards,
\citet{pou2004a,pou2004b}
carried on a systematic
study of the spectral
variability in this source,
using five 
$\sim$15 ks long observations,
that were taken
during one year
with a time spacing of $\sim$3 months.
These authors concluded that
the spectral variability
is dominated by an 
emerging/disappearing
steep power-law component,
which is in turn modified
by a slightly ionized
variable absorber.
The fitted absorber becomes
more ionized and less
opaque as the continuum
flux increases, supporting
the idea that a fraction
of the soft X-ray emission
may be due  to re-emission
of the absorbed continuum
in an extended region of
photoionized gas.
%
%
%
An alternative explanation
of the same \th \xmm \th \th
datasets was however
readily proposed in
the framework of
blurred reflection model
\citep[][hereafter F05]{fab2005}.
This model prescribes
that AGN spectral variability
is due to the degree of light-bending as
the primary power law emitting spot
moves in a region of stronger gravity.
Low flat states, such as the ones
observed in \th \source,
are  extreme reflection dominated cases
occurring when the
primary emission is almost 
completely focused down to
the disk and do not reach
the observer.
The broader spectral
coverage provided
by two subsequent $Suzaku$ observations
of \th \source \th \th
did not break this models degeneracy.
The variable excess observed above
$\sim$15 keV
can be either explained by
reflection \citep{wal2010, pal2013} or
by reprocessing of the
primary emission in
a partially covering,
Compton-thick, 
screen of gas
\citep{tur2009}.
The high ionization
parameter suggests
that this absorber
may be part of
a clumpy disk wind
located
within the broad line
region (BLR).
\\ 
In this paper
we present the
longest exposure
EPIC-pn dataset
of 1H0419-577
obtained so far.
The \th \xmm \th \th
observation was taken
simultaneously to
a HST-COS observation
in the UV \citep{edm2011},
and caught the source in
an intermediate flux
state (Fig. \ref{fluxes.fig}).
In the Reflection Grating
Spectrometer (RGS) spectrum of
this dataset
\citep[already presented in][hereafter Paper I]{dig2013}
we detected
a lowly ionized
absorbing gas
\citep[also observed in $Suzaku$,][]{win2012}.
We found that the X-ray
and the UV
absorbing gas
\citep{edm2011}
are consistent
to be one and the same.
The low gas density estimated
in the UV together with
the low
ionization parameter
that we measured
in the X-ray imply
a galactic scale location
for the absorbing gas
($d \sim 4 \, \rm kpc$).
In this respect, the
warm absorber in \th
\source \th \th represents
a unique case,
being the first X-ray absorber
ever detected so far
away from the nucleus.
The absorbing gas
does
not have an emission
counterpart as
more highly ionized
lines, produced by e.g.
\ion{O}{vii}, \ion{O}{viii}, and
\ion {Ne}{ix}, are
the most prominent
emission features in
the X-ray spectrum.
The photoionization modeling
of the  X-ray and UV
narrow emission-lines 
confirmed that 
they are produced
by a more highly ionized
gas phase, located
closer ($\sim$1 pc)
to the nucleus.\\ 
In the present analysis
we exploit the simultaneous
UV and optical 
(thanks to the \th \xmm \th \th
Optical Monitor) 
coverage to model 
the X-ray
spectrum in a broadband
context
using Comptonization.
The paper is organized as follows:
in Sect. \ref{obs} 
we explain the data reduction procedure;
in Sect. \ref{spec} 
we present the spectral analysis of our dataset;
in Sect. \ref{var}
we apply our best fit model to the past \th \xmm \th \th
datasets, with the aim of explaining the historical spectral variability; 
finally in Sect. \ref{disc} we
discuss our results and
in Sect. \ref{conc} we outline our conclusions.
The cosmological parameters used are:
\ho=70 \kmsmpc, \omegam=0.3, \omegalambda=0.7.
The
C-statistics \citep{cas1979} is used
throughout the paper,
and errors are quoted at 90\% confidence levels
($\Delta C=2.7$). In all the
spectral models presented in the
following we consider the Galactic 
hydrogen column density
from \citet[][\nh=$1.26 \times 10^{20}$ \colc]{kal2005}.


\begin{table}[t]
\caption{OM, FUSE and HST/COS continuum values for 1H0419-577.}
\centering
\begin{tabular}{lcccccc}
\hline \hline
Instrument & $\lambda$ 
\tablefootmark{a} 
& 
$\Delta \lambda$
\tablefootmark{b} 
& 
$ F_{\lambda}$ 
\tablefootmark{c}
& 
$\frac{\Delta F_{\lambda}}{F_{\lambda}}$ 
\tablefootmark{d}
& 
Norm
\tablefootmark{e}
\\
 & \AA & \AA & \ergsca &  & \\
\hline
OM-b    &  4340   & 1307 &  $3.37 \times 10^{-15} $   &  0.2\%  & $2.9_{-0.7}^{+0.7}$ \\  
OM-uvw1 &  2910   & 1829 &  $1.10 \times 10^{-14} $   &  0.3\%  & $3.2_{-1.1}^{+1.6}$ \\  
OM-uvm2 &  2310   & 1333 &  $1.32 \times 10^{-14} $   &  0.6\%  & $4.5_{-0.9}^{+1.8}$ \\  
OM-uvw2 &  2120   & 811  &  $1.40 \times 10^{-14} $   &   0.9\% & $3.1_{-1.4}^{+2.5}$\\  
COS     &  1500   & 10   &  $2.44 \times 10^{-14}$    &  4\%    & $2.2_{-0.7}^{+1.2} $\\  
FUSE    &  1031   & 10   &  $2.80   \times 10^{-14}$  &  14\%   & $1.1_{-0.3}^{+1.8}$ \\  
FUSE    &  1110   & 10   &  $3.66   \times 10^{-14}$  &  16\%   & $1.7_{-0.4}^{+2.2}$ \\  
\hline
\end{tabular}
 \tablefoot{
 \tablefoottext{a}{Centroid of the spectral bin for each instrument.}
 \tablefoottext{b}{Spectral bin width.
  For the OM-filters, we used the filters width.
  We instead assigned a narrow spectral width of 10 \AA\ to the COS and FUSE flux measurements.}
 \tablefoottext{c}{Flux density.}
 \tablefoottext{d}{Statistical error for the flux density.}
\tablefoottext{e}{Intercalibration factor relative to the EPIC-pn,  with errors.}
 }
\label{fluxes.tab}
\end{table}
%

\section{Observations and data preparation}
\label{obs}

\source \th \th was observed in May 2010
with \th \xmm \th \th for $\sim$167 ks.
The observing time was split into two
observations (Obs. ID 0604720301 and
0604720401 respectively) which were performed in 
two consecutive satellite orbits. 
For the present analysis, we used
the EPIC-pn \citep{stru2001} and the 
Optical Monitor \citep[OM,][]{mas2001} data.
In the UV, besides the
HST-COS observation
simultaneous to our
\th \xmm \th \th observation,
the source has been
observed twice with
the Far Ultraviolet
Space Explorer (FUSE),
respectively in 2003
and in 2006. In this
analysis we used
the FUSE flux measurements
reported in the literature
\citep{dun2008,wak2009}.
Finally, we retrieved all the
available archival datasets from
the \th \xmm \th \th archive and we
used them to study the
source variability.

\subsection{The X-ray data}
\label{x.dat}
We processed the present datasets and
all the archival Observation
Data Files (ODF) with the \th \xmm \th \th Science
Analysis System (SAS), version 10.0,
and with the HEAsoft FTOOLS, version 6.12.
We refer the reader to Paper I for
a detailed description of the data reduction.
\\
For the present datasets,
we extracted the EPIC-pn spectra from both
0604720301 and 0604720401 observation.
We checked the stability of the spectrum
in the two observations and we found no
flux variability larger than $\sim$ 7\%. 
Therefore, we
summed up the two spectra into a single
combined spectrum with a net exposure
time of $\sim$ 97 ks after the background
filtering. We used
the FTOOLS \textsf{mathpha} and \textsf{addarf}
to combine respectively the spectra
and the Ancillary Response Files (ARF).\\ 
We reduced all the archival datasets
following the same standard procedure
described in Paper I, and
we discarded the datasets 
with ID 0148000301 and 0148000701 because
they show a high 
contamination by background flares.
Hence, we created the EPIC-pn spectra and
spectral response matrices for all the
good datasets.\\
We fitted all the X-ray spectra
in the 0.3--10 keV band and we
rebinned them in order to have
at least 20 counts 
in each spectral bin,
although this is not strictly
necessary when using the
$C$ statistics.
In Table \ref{archive.tab} we provide
the most relevant information of each
\th \xmm \th \th observation and we
label them with numbers, 
following a chronological
order. 

\subsection{The optical and UV data}
\label{opt.dat}
As also described
in Paper I, in our \th \xmm \th \th observation
OM data were collected in 4 broad-band
filters: B, UVW1, UVM2, and UVW2.
In the present analysis we used 
the OM filters count-rates for the
purpose of spectral fitting. Therefore,
we also retrieved from the ESA
website\footnote{
\url{http://xmm2.esac.esa.int/external/xmm_sw_cal/calib/om_files.shtml}} 
the spectral response matrices 
correspondent to each filter.
We corrected the flux in the B filter
to account
for the host galaxy
starlight contribution.
For this, we used the same correction
factor (56\%) estimated in M11
for the stellar 
bulge of Mrk 509. Indeed,
since Mrk 509 hosts a BH with a mass
similar to the one in \th \source \th \th,
also the stellar mass of the
bulge should be similar in
this two galaxies \citep[e.g.,][]{mer2001}.
\\
In Paper I we derived a value for the
UV flux of the source at 1500 \AA\
from the continuum of the 
HST-COS spectrum.
Moreover two other UV fluxes
measured with FUSE are reported
in the literature,
at
1031 \AA\ \citep{wak2009} 
and 1110 \AA\ \citep{dun2008},
respectively.
When the overlapping wavelength
region between COS
(2010 observation) and FUSE (2003 and
2009 observations) is
considered,
the level of the UV
continuum of the source 
is the same
\citep[see][]{wak2009,edm2011}.
Therefore,
we could safely
fit the FUSE fluxes
together
with the COS, OM and the EPIC-pn data
that were simultaneously 
taken in 2010.
For this purpose,
we converted
the UV fluxes back to count rates.
We used the HST-COS sensitivity curve and
the FUSE effective area (see also M11)
for this.
We outline all the UV and optical
continuum values for \th \source \th \th
in Table \ref{fluxes.tab}.
\\
To check for a possible variability
of the source in the optical-UV,
we obtained also the OM fluxes from
the archival \th \xmm \th \th observations.
For all the archival datasets,
except Obs.1, OM data were
available in the U, B, V, UVW1, and UVW2 
filters.

\begin{figure}[t]
\centering
\includegraphics[angle=-90,width=0.5\textwidth]{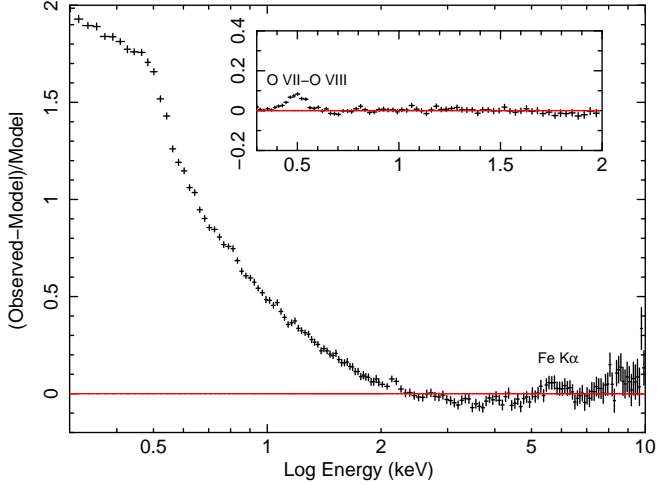}
\caption{\textit{Main Panel}: model residuals to a power law model, in
the 0.3--10.0 keV band. 
A prominent soft excess in the $\sim$ 0.3--2.0 keV band, and
a shallow excess at the rest frame energy of the \th \fek \th transition are seen.
\textit{Secondary Panel}: model residuals to the broadband phenomenological model, in 0.3--2.0 keV band. 
An arc-shaped feature at the
rest frame energy of the \ion{O}{vii}-\ion{O}{viii} transitions is seen.}
\label{res.fig}
\end{figure}


\section{Spectral modeling}
\label{spec}

\subsection{A phenomenological model}
\label{fen}

We started with a pure phenomenological
modeling of the present EPIC-pn spectrum,
using SPEX \citep{kaa1996}, version 2.04.00.
We first attempted to fit the spectrum
in the 2.0--10.0 keV energy region
with a canonical simple power law
($\Gamma \sim 1.6$).
The broadband residuals
(Fig. \ref{res.fig})
show large deviations
from this simple model.
Indeed, besides a prominent
soft excess in the
0.3--2.0 keV band,
the model
does not account for a broad
trough between 2.0 and 4.0 keV.
To phenomenologically
account for this
nontrivial spectral
shape,
a combination of 4
different spectral
slopes would be required
all over the 0.3--10.0 keV band.
Nonetheless,
a prominent peak
in the model residuals
(Fig. \ref{res.fig}),
at $\sim$ 0.5 keV,
is still unaccounted.
We identified this feature as due
to the blend of
the \ion{O}{vii}-\ion{O}{viii}
lines that
we detected in the
simultaneous RGS spectrum (Paper I).
Furthermore, a shallow excess
is seen at $\sim$ 5.5 keV.
Fitting this feature with
a delta-shaped emission line,
centered at the nominal
rest frame energy of the
\th \fek \th line-transition,
does not leave any prominent
structure in the residuals.
If the line width
is left free to vary, the
fitted line-width 
($\sigma=300\pm200$ eV)
is well consistent with what
previously reported
for this source
\citep{tur2009,pou2004a,pou2004b}.
We attempted also to decompose
the \th \fek \th in a
combination of a
broad plus a narrow
component, but
this exercise 
did not lead
to a conclusive result.
Despite the good
data quality of the present
dataset,
the line-width
of the broad component 
and the
normalization of the narrow
component cannot be 
constrained
simultaneously.\\
In conclusion, the long
exposure time of the present
EPIC-pn spectrum unveiled
a complex continuum
spectral structure, which
calls for a more 
physically motivated
modeling to be fully understood.

\subsection{Reflection Fitting}
\label{refl}

\begin{table}[t]
\caption{Best fit parameters for the reflection fitting.}
\begin{center}
\begin{tabular}{lc}
\hline
\hline
 \multicolumn{2}{c}{Simple reflection model} \tablefootmark{a}\\
 \hline
 $\Gamma$ \tablefootmark{b}& $2.17 \pm 0.01$  \\ 
 $R_{\rm in} $\tablefootmark{c}   &  $1.6 \pm 0.1$ $R_{\rm g}$ \\ 
 $R_{\rm break}$\tablefootmark{d} &  $ 7 \pm 2$ $R_{\rm g}$ \\ 
 $q_{\rm in}$ \tablefootmark{e}     &  $5.9 \pm 0.3$ \\ 
 $q_{\rm out} $\tablefootmark{f}     &  $2.6 \pm 0.2$ \\ 
 Incl.  \tablefootmark{g}    &  $20 \pm 8 $ \\          
 $\xi $  \tablefootmark{h}   &  $20.8 \pm 0.3$ \\
 $F_{\rm pow} $ \tablefootmark{i}  &  $16.6 \pm 0.07 \times 10^{-12} \ergsc $\\
 $F_{\rm ref} $  \tablefootmark{l} &  $7.4_{-0.08}^{+0.47} \times 10^{-12} \ergsc $\\
 $\rm C/d.o.f $    & 383/240\\ 
 \hline
 \multicolumn{2}{c}{Composite disk model}\tablefootmark{m}\\
 \hline
 $\Gamma$ \tablefootmark{b}  & $2.13 \pm 0.01$  \\ 
 $R_{\rm in, \,2}=R_{\rm out,\,1}$ \tablefootmark{n}  &  $1.90 \pm 0.06$ $R_{\rm g}$ \\ 
  Incl    \tablefootmark{g}  &  $41 \pm 2 $ \\ 
 $q_{\rm in}$ \tablefootmark{o}    &  $6 \pm 2 $ \\ 
 $\xi_{\rm in} $  \tablefootmark{p}   &  $89 \pm 11 $ \\ 
 $q_{\rm out} $ \tablefootmark{q}  &  $3.7 \pm 0.2 $ \\ 
 $\xi_{\rm out} $  \tablefootmark{r}   &  $20 \pm 1 $ \\ 
 $F_{\rm pow} $  \tablefootmark{i} &  $15.9 \pm 0.1 \times 10^{-12} \ergsc $\\
 $F_{\rm ref} $ \tablefootmark{l}  &  $8.0 \pm 0.1 \times 10^{-12} \ergsc $\\
 $\rm C/d.o.f $    & 354/240\\
 \hline
\end{tabular}
\end{center}
\label{refl.tab}
\tablefoot{
\tablefoottext{a}{Xspec syntax: PHABS*(POW+KDBLUR2(REFLIONX)).}
\tablefoottext{b}{Photon index.}
\tablefoottext{c}{Disk inner radius.}
\tablefoottext{d}{Break radius above which the emissivity profile of the disk changes slope.}
\tablefoottext{e}{Index of the emissivity profile of the disk for $r \leq R_{\rm break}$.}
\tablefoottext{f}{Index of the emissivity profile of the disk for $r \geq R_{\rm break}$.}
\tablefoottext{g}{Disk inclination.}
\tablefoottext{h}{Disk ionization parameter.}
\tablefoottext{i}{Unabsorbed flux of the main power law in the 0.5--10 keV band.}
\tablefoottext{l}{Unabsorbed flux of the reflected power law in the 0.5--10 keV band.}
\tablefoottext{m}{Xspec syntax: PHABS*(POW+KDBLUR(REFLIONX)+KDBLUR(REFLIONX)).}
\tablefoottext{n}{Inner radius of the second reflection, 
that is coupled with the outer radius of the first one.}
\tablefoottext{o}{Index of the disk emissivity profile for the first reflection.}
\tablefoottext{p}{Disk ionization parameter for the first reflection.}
\tablefoottext{q}{Index of the disk emissivity profile for the second reflection.}
\tablefoottext{r}{Disk ionization parameter for the second reflection.}}
\end{table}

\begin{figure}[h]
\centering
\includegraphics[angle=90,width=0.5\textwidth,]{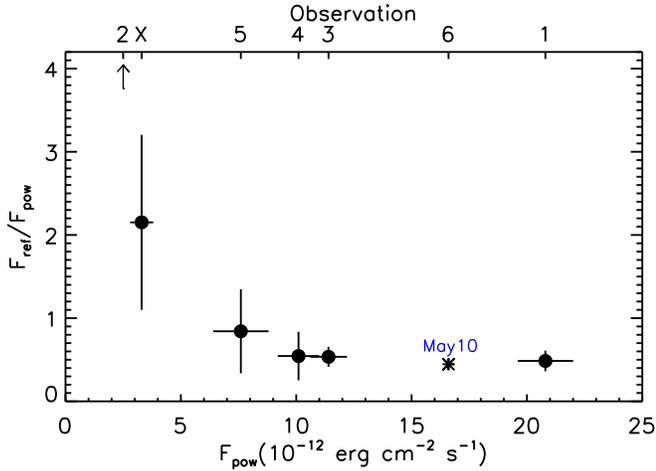}
\caption{
Reflection fraction as a function
of the power law flux, with errors.  
The reflection fraction is given
by the ratio between the 0.5--10.0 keV fluxes 
of the reflected
component to the primary power law.
The value for the
present dataset is labeled, while all the others
data points, including a lower limit
indicated by an arrow, are taken from F05.
The observation numbers for each data point
are labeled on the upper axes (see Table \ref{archive.tab}). 
Note that
the data point labeled with a ''X''  has not
been used in the present analysis 
(see Sect. \ref{x.dat}).
}
\label{refl.fig}
\end{figure}


At first, we tested
a disk reflection
scenario for the present
spectrum.
As noticed
in Sect. \ref{intro}, this
model has been already
successfully applied
to Obs. 1--5 (F05).
Besides the main
power law continuum,
the second relevant
spectral component in this
model is a relativistically
smeared reflected power
law, which is 
thought to be 
produced in an ionized
accretion disk.\\ 
We fitted the spectrum
with Xspec \citep{arn1996} 
version 12.0,
and we used PHABS
to account for
the Galactic hydrogen column
density along the line of sight.
We used REFLIONX
\citep{ros2005}
to model the reflected
component, and we
left the ionization
parameter of the reflector
free to vary. Hence,
we accounted for the relativistic
effects from an accretion
disk surrounding a rotating black hole
\citep{lao1991}
with KDBLUR2. 
The free parameters in
this component
are the disk inclination
and inner radius, 
along with the slopes and the
break radius of the broken power law
shaped emissivity profile. 
We kept the outer
radius frozen to the
default value of
400 gravitational radii
($R_{\rm g}$),  and we set
the iron abundance to
the solar value \citep{and1989}. 
We extended the model
calculation to a larger energy range
(0.1--40 keV) to avoid
spurious effect due to a truncated
convolution.
We also attempted to fit
the spectrum with a
composite disk model (see F05), 
splitting the disk in two regions
with different ionization,
to mimic a more realistic
scenario
where the disk ionization
parameter varies with the radius.
However, with a simple 
reflection model
we already obtained 
a statistically good fit,
that was not strikingly
improved ($\Delta C=29$)
using
a more complex composite
disk model.
We list the best fit
parameters of the reflection fitting
in Table \ref{refl.tab}.
Overall,
our result agrees  with 
the main predictions 
of the physical
picture proposed in F05. 
The black hole hosted in \th \source \th \th
may be rapidly spinning as
suggested by the proximity
of the fitted disk inner radius
to the value of
the innermost stable orbit of a 
maximally rotating Kerr black hole.
The steep emissivity profile of the
disk indicates that it is illuminated
mostly in its inner part, as it is 
expected
if the primary continuum is emitted
very close to the BH. In this framework,
the historical source variability
is due to the variable light bending, which
may produce a negative trend of the reflection
fraction with the power law flux.
Our results are consistent with the
general trend noticed
in F05 (Fig. \ref{refl.fig}).

\subsection{Broadband spectral modeling}
\label{opt.mod}

\begin{figure*}[t]
\centering
\includegraphics[angle=90,width=\textwidth,]{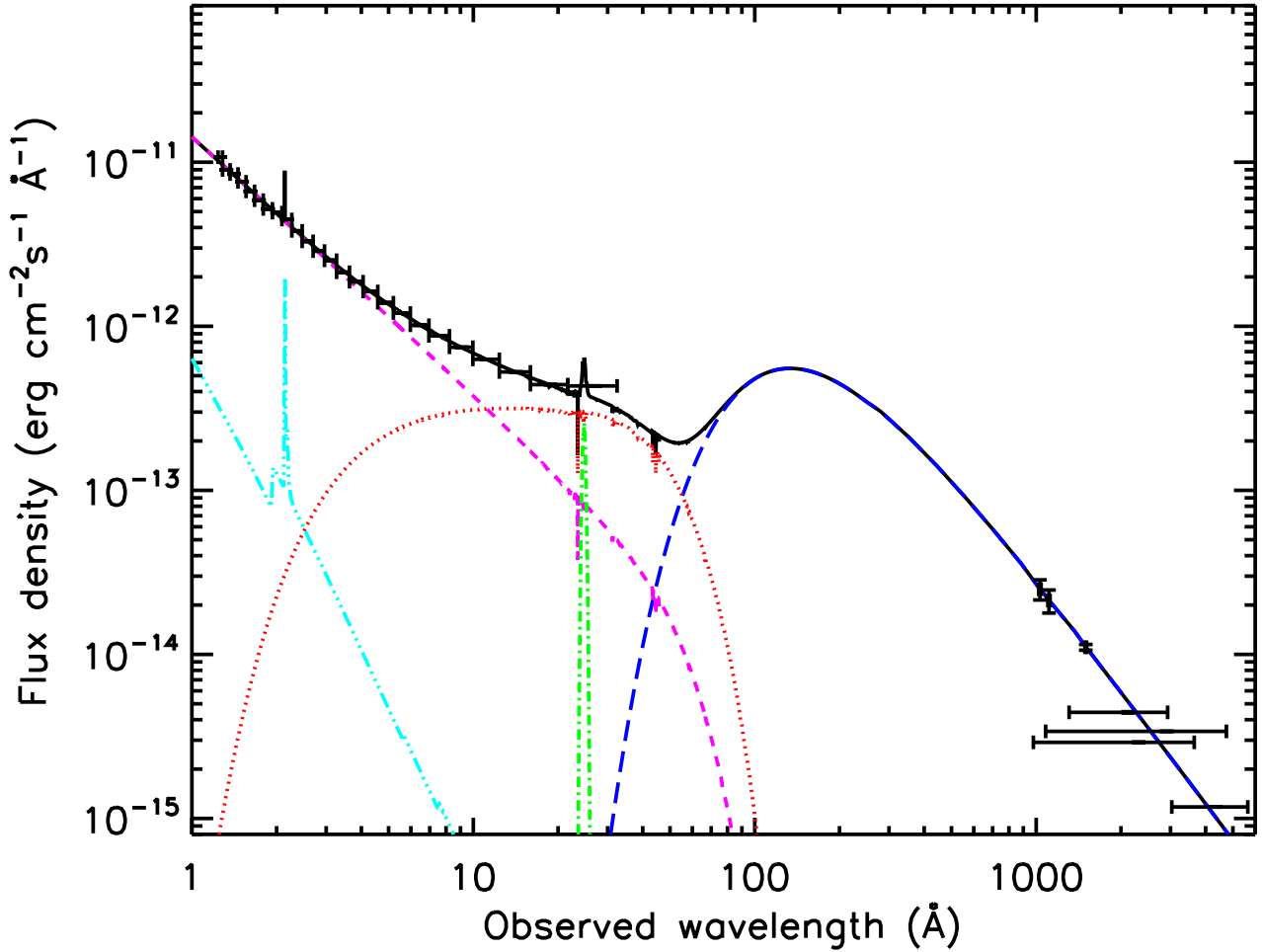}
\caption{Best fit comptonization model for \th \source. 
Solid line: total model.
Crosses: fluxed EPIC-pn spectrum (rebinned for clarity),
OM, COS and FUSE data points, with errors.
Long dash line: disk blackbody component.
Dot line: warm Comptonized component.
Dash line: hot Comptonized component.
Dash dot line: gaussian \ion{O}{vii} (triplet) emission line.
Dash dot dot line: cold reflection component, accounting for the \th \fek \th emission lines.}
\label{fitcomt.fig}
\end{figure*}


The AGN emission can
be also produced by thermal
Comptonization 
(see Sect. \ref{intro}).
This model has the advantage
of explaining
AGN emission
in a consistent way
over the entire 
optical, UV, and X-ray energy range
(e.g. Mrk 509, M11, P13).
Indeed,
the disk
blackbody temperature
that can be constrained
from a fit of the
optical/UV data
serves as input
for the Comptonized
components that
produce the
X-ray continuum.
The model includes
both a warm (hereafter labeled as ''wc``) 
and a hot Comptonizing corona (hereafter labeled as ''hc``)
to cover the entire X-ray
bandpass.
Given the simultaneous 
X-ray, UV and optical
coverage available in the present
case, it is worthwhile testing
also this scenario.\\
We fitted the EPIC-pn spectrum of \th \source \th
together with the COS, FUSE and OM count-rates
with SPEX.
We left the
normalization of each instrument relative
to the EPIC-pn as a free
parameter to account for
the diverse collecting
area of different detectors. 
In the fit we both accounted
for the Galactic
absorption and for 
the local warm absorber that we detected
in Paper I. For the former,
we used the SPEX collisionally-ionized plasma
model (HOT), setting a
low temperature (0.5 eV) 
to mimic a neutral gas.
The cosmological redshift
(z=0.104) was also considered
in the fit. The final multicomponent
model is plotted in Fig. \ref{fitcomt.fig}.\\
%
%
We used the disk-blackbody
model (DBB) in SPEX to model
the optical-UV emission
of \th \source. This model is based
on a geometrically thin, optically thick, 
Shakura-Sunyaev accretion disk \citep{sha1973}. 
The DBB spectral shape 
results from the weighed sum 
of the different blackbody spectra
emitted by annuli of the disk
located at different radii. 
The free parameters
are the maximum temperature in the
disk ($ T_{\rm max}$) and the normalization
$ A=R^2_{\rm in} \cos i$, where $R_{\rm in}$
is the inner radius of the disk
and $i$ is the disk inclination.
We kept instead the ratio
between the outer and the inner
radius of the disk frozen to
the default value of $10^3$.
The parameters of the
disk-blackbody best fitting
the data (Fig. \ref{fitcomt.fig}, long dashed line)
are: 
$T_{\rm max}=56 \pm 6 \rm \,eV$ 
and 
$A=(1.2 \pm 0.6) \times 10^{26} \, \rm cm^{2}$.
The fitted
intercalibration factors
between OM, COS and FUSE
and EPIC-pn,
with errors,
are reported
in Table \ref{fluxes.tab}.
The effect
of these intercalibration
corrections is within
the errors of the disk
blackbody parameters 
given above.\\
We used the COMT model
in SPEX, which is based on the
Comptonization model of \citet{tit1994},
to model the X-ray continuum.
The seed photons in this model have a
Wien-law spectrum with temperature
$T_0$. In the fit we coupled
$T_0$ to the disk temperature
$T_{\rm max}$. The other free
parameters are the electron temperature
$T$ and the optical depth $\tau$ of the Comptonizing
plasma. A combination of two Comptonizing
components fits the entire EPIC-pn 
spectrum. The warm corona ($T_{\rm wc} \sim 0.7$ keV) 
is optically thick ($\tau_{\rm wc} \sim 7$)
and produces the softer part of the X-ray
continuum, below $\sim$2.0 keV
(Fig. \ref{fitcomt.fig}, dotted line).
On the other hand, the hot corona 
($T_{\rm hc} \sim 160 $ keV) is optically
thin ($\tau_{\rm hc} \sim 0.5 $) 
and accounts for the X-ray emission
above $\sim$2 keV
(Fig. \ref{fitcomt.fig}, dashed line).\\
Hence,
we identified
the remaining features 
in the model residuals
as due to the
\ion{O}{vii}--\ion{O}{viii}
(Fig. \ref{fitcomt.fig}, dash dot line)
and \th \fek \th emission
lines
(Fig. \ref{fitcomt.fig}, dash dot dot line).
We added to the fit
a broadened Gaussian line, with
the line-centroid and the line-width
frozen
to the values that we
obtained in the RGS fit (Paper I)
to account for the \ion{O}{vii}
emission. The fitted
line luminosity is consistent
with what reported in Paper I.
The shallower
\ion{O}{viii} line that was present
in the RGS spectrum
is instead
undetected in the EPIC data.
In a Comptonization framework,
a possible origin for
the \th \fek \th emission
is reflection from a
cold, distant matter
(e.g., from the torus).
We have shown
in Sect. \ref{fen} 
that the \th \fek \th
line in \th \source \th 
might also be broad.
Detailed study of
the properties of the
\th \fek \th emission line
produced in cold matter
show that
in some conditions
the line may appear
broadened because of
the blend between the
main line core and
the so-called "Compton shoulder"
\citep[see e.g.,][]{yaq2011}.
The predicted apparent
line broadening is consistent
with what we have obtained
in Sect \ref{fen}
from a phenomenological
fit of a possible
line-width.
We added a REFL component
to the fit to test this
possibility.
We considered
an incident power law with
a cutoff energy of 150 keV,
and with the same slope and normalization
that we derived from the phenomenological fit
(Sect. \ref{fen}).
We set a null ionization
parameter and a low gas
temperature ($T \sim 1 \rm \,eV$)
to mimic a neutral reflector
and, to adapt
the model to the data,
we left only the scaling factor
\footnote{In the REFL 
model, 
the total spectrum $N(E)$
is given by:
$N(E)=N_{i}(E) + s R(E)$
where:
$N_{i}(E)$ is the incoming spectrum,
$R(E)$ is the reflected spectrum, and
$s$ is the scaling factor.
See the SPEX manual for details.
} ($s$)
free to vary. A reflected
component with $s=0.3 \pm 0.1$ 
satisfactorily fits the \th \fek \th line
and slightly adjusts 
the underlying continuum
($\Delta C=-7$). \\
We list the luminosities of all
the model components in Table
\ref{fitcomt.tab} while the parameters
and errors of the Comptonized components
are outlined in Table \ref{var.tab}. 

\begin{table}[h]
\caption{Luminosities of the broadband model components.}
\centering
\begin{tabular}{lc}
\hline \hline
Model component
& 
$\log L $
\tablefootmark{a}
\\
&
\ergs\\
\hline
Disk blackbody & 45.70 \\
Warm corona & 44.79 \\
\ion{O}{vii} emission line & 42.83 \\
Hot corona & 45.64 \\
Cold reflection & 43.78 \\
\hline
\end{tabular}
\tablefoot{
\tablefoottext{a}{Total intrinsic luminosity.}
}
\label{fitcomt.tab}
\end{table}


\begin{figure}[h]
\centering
\includegraphics[angle=90,width=0.5\textwidth,]{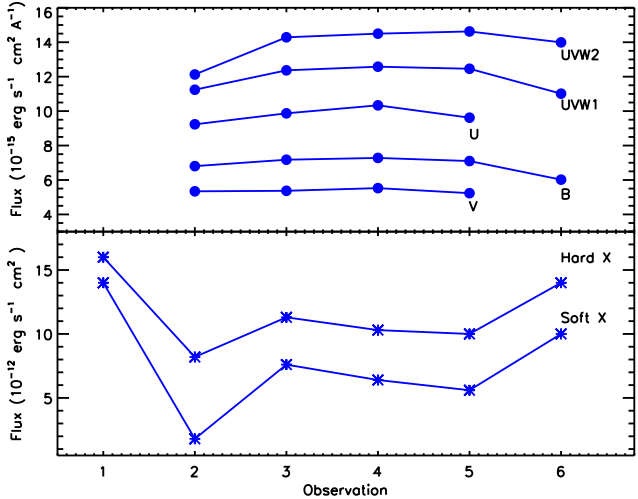}
\caption{
\textit{Upper panel}: variability of the optical
and UV fluxes in the OM broadband filters, 
for the archival and present \th \xmm \th \th datasets.
\textit{Lower panel}: variability
of the soft (0.5--2.0 keV) and
hard (2.0--10.0 keV) X-ray fluxes.}
\label{var_fl.fig}
\end{figure}


\begin{figure}[h]
\centering
\includegraphics[angle=90,width=0.5\textwidth,]{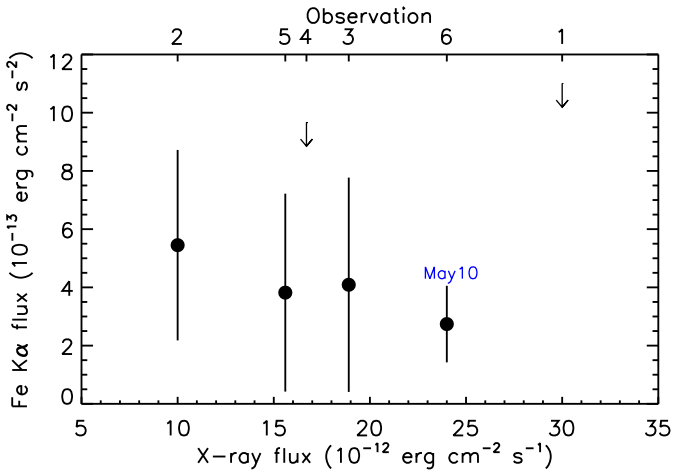}
\caption{Historical and present observed fluxes of the \th \fek \th emission line,
with errors, as a function of the source flux in the 0.5--10.0 keV band. The arrows
represent upper limits for the line flux. The value for
the present dataset is labeled and the observation numbers for each data point 
(see Table \ref{archive.tab})
are labeled as well on the upper axes.}
\label{var_fe.fig}
\end{figure}


\section{The historical spectral variability}
\label{var}

\subsection{The baseline model}

In Fig. \ref{var_fl.fig} we plot
the historical X-ray, optical and
UV fluxes of \th \source, from
the archival \th \xmm \th \th observations
and from the present dataset.
The optical-UV flux
of the source has been stable
throughout the $\sim$8 years
spanned by the available 
OM observations (Obs 2--6).
Nevertheless, as already pointed
out in Sect. \ref{sou}, in the
soft X-rays the source has been observed
by \th \xmm \th \th in a variety of
flux states 
(see Fig. \ref{var_fl.fig}), 
ranging from the
deep flux minimum of September 2002
(Obs.2) to the highest state of
December 2000 (Obs.1).
We used
the Comptonization
model that successfully fitted
the present dataset as a
baseline model for the fit of
the past \th \xmm \th \th
observations.
Because the maximum observed
variability in the optical-UV
($\sim20\%$)
is within the errors in the
disk blackbody parameters
that we derived in the broadband
fitting (Sect. \ref{opt.mod}),
we assumed
the same seed photons temperature of
the present dataset in
the baseline model.\\
In Fig. \ref{var_fe.fig} 
we plot the historical fluxes
of the \th \fek \th emission line,
that were measured from
a phenomenological fit 
of the archival and present
\th \xmm \th \th datasets
in the 2.0--10.0 keV band.
Although
the line is not well constrained
in any of the archival 
datasets, its flux is
however consistent to have
been stable in the 
$\sim$ 10 years long
period covered by \th 
\xmm \th \th observations.
Therefore, also
the cold reflection
continuum associated
to the \th \fek \th
line should have 
remained constant.
Since we are mainly
interested in studying
the source variability in
the soft X-ray band,
we included in the
baseline model
just a delta function
to account
for the \th \fek \th
emission line.
Indeed,
the addition
of a cold reflection continuum
is not critical for the 
resulting parameters.
Finally, we included 
unresolved  \ion{O}{vii}-f
and \ion{O}{viii}-\lia \th 
emission lines in the
baseline model.
Indeed, 
previous analysis of
the RGS spectrum
\citep{pou2004b} 
has shown that
\ion{O}{vii} and
\ion{O}{viii}
lines 
were present
in Obs 2--5.
The baseline model
provides
a formally acceptable
fit for all the
datasets except
Obs. 2, namely
the lowest flux
state.

\subsection{The low flux state}

\begin{figure*}[t]
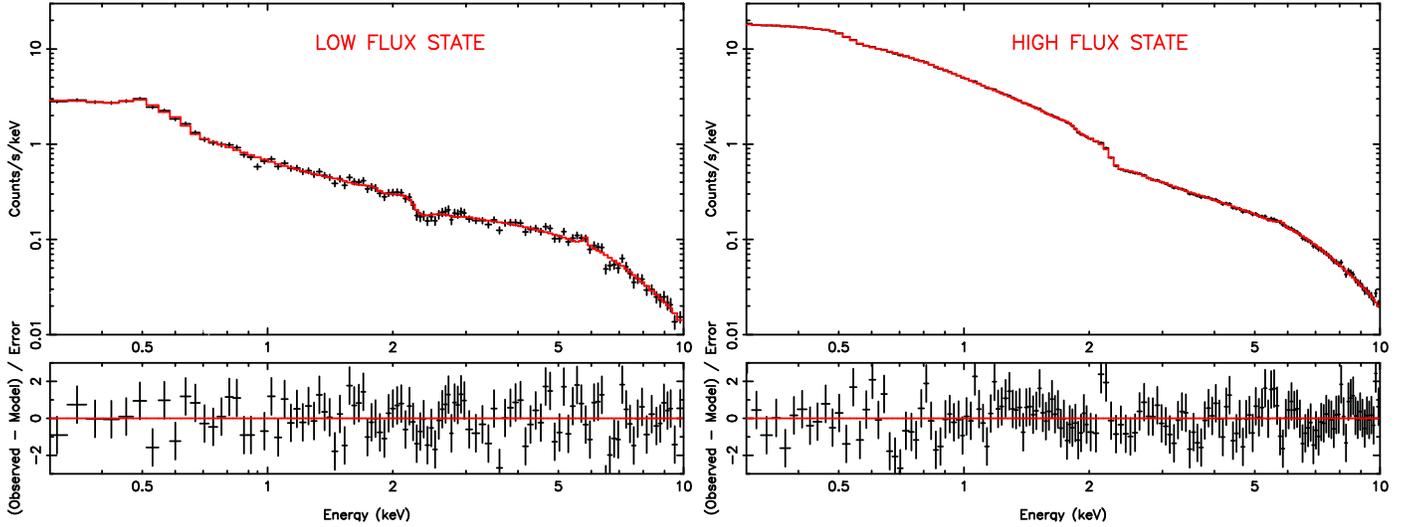

  \centering
  \begin{minipage}[c]{1.0\textwidth}
    \includegraphics[angle=-90,width=.50\textwidth]{lowstate.ps}
    \includegraphics[angle=-90,width=.50\textwidth]{mystate.ps}
  \end{minipage}
  \caption{
EPIC-pn spectrum of \th \source \th \th in the 
low flux state (Obs.2, \textit{left panel}),
and in the present intermediate flux state (Obs.6, \textit{right panel}).
The best fit Comptonization model is displayed as a solid line. 
Model residuals, in terms of $\sigma$, are also shown.
We rebinned the spectra for clarity purposes.
Note that for the present datasets, 
error bars are as large as the thickness of the model line.
}
\label{lowstate.fig}
\end{figure*}

We show
in Fig. \ref{lowstate.fig}
a comparison between the
low-state spectrum and
the present dataset.
At a first glance,
the spectrum
appears much flatter in
the $\sim$1.0-2.0 keV
band and displays 
a peak at $\sim$0.5 keV
resembling
the shape of an emission line. 
Indeed,
this feature is well
modeled by
the
\ion{O}{vii}-\ion{O}{viii} emission lines
that are present in the
baseline model.
The fitted line fluxes
($\rm F_{\ion{O}{vii}}=(16 \pm 8)\times 10^{-5}$ \th \phsc \th
and
$\rm F_{\ion{O}{viii}}=(8 \pm 4)\times  10^{-5}$ \th \phsc)
are consistent
with what previously reported
for the RGS spectrum of
this dataset \citep{pou2004a}.
The residuals to
the baseline model
display
a deep trough,
approximately in same
energy region where
the spectrum flattens
out with respect to the
present dataset.
We added to the fit
a neutral absorber
located at the redshift of the source,
modeled by a cold (T=0.5 eV)
collisionally-ionized 
plasma (HOT model in SPEX)
to attempt adapting
the baseline model
to the low-state
spectrum.
By letting all the Comptonized emission
being absorbed by 
a partially covering
($\sim$ 60\%),
thick
(\nh $\sim 5 \times 10^{22}$ \th \colc) 
cold gas,
we achieved
a statistically good
(C/d.o.f.=163/152)
fit to the data,
accounting for
all the features left
in the residuals
by an unabsorbed
model. 

\begin{table*}[t]
\caption{Best fit parameters and errors, 
for the absorbed Comptonization model.}
\centering
\begin{tabular}{lccccccccccc}
\hline
\hline
Obs &
$T_{\rm wc}$ \tablefootmark{a}&
$\tau_{\rm wc}$ \tablefootmark{b}&
$\rm F^{wc}_{0.5-10.0\,keV}$ \tablefootmark{c}&
$T_{\rm hc}$ \tablefootmark{a}&
$\tau_{\rm hc}$\tablefootmark{b} &
$\rm F^{hc}_{0.5-10.0\,keV}$ \tablefootmark{c} &
$\rm C/d.o.f.$\tablefootmark{d} & 
\nh \tablefootmark{e}& $C_{\rm V}$\tablefootmark{f} & 
$\rm \Delta C$\tablefootmark{g}\\
& keV & & $10^{-12}$\ergsc & 
keV & & $10^{-12}$ \ergsc &  
& $10^{21}$ \colc & &\\
\hline
1 
& $ 1.1 \pm 0.2$ & $6.5 \pm 0.9$ & $5.8 \pm 0.5$ 
& $140 \pm 30$ & $ 0.2 \pm 0.1$ & $24 \pm1$ 
& 157/141 & $ \leq 1 $ & 0.5 (f) & -7 \\
2 
& $ 0.6 \pm 0.1$  & $ 5.3 \pm 0.6 $ & $0.45 \pm 0.05 $ & 
$150^{+80}_{-40} $   &$ 0.7 \pm 0.3$ & $9.1 \pm 0.6 $
& 314/151 & $47 \pm 10$ & $0.6 \pm 0.1$ & -143\\
3 
& $ 0.7 \pm 0.1 $  & $6.3 \pm 0.5 $ & $4.6 \pm 0.8$ 
& $160 \pm 60$ & $ 0.3 \pm 0.2 $ & $14.3 \pm 0.3$ 
& 187/151 & $7 \pm 3$ & $0.5 \pm 0.1$ & -26\\
4 
& $ 0.67 \pm 0.09$ & $ 7.6 \pm 0.7$  & $4.3 \pm 0.9$  
& $130^{+70}_{-40} $ & $0.6^{+2}_{-0.2}$ & $13 \pm 1$ 
& 158/139 & $9 \pm 4$ & $0.5 \pm 0.2$ & -14\\
5 
& $ 0.46 \pm 0.05$ & $9.0 \pm 0.7$  & $2.9 \pm 0.5$ 
& $170 \pm 40 $ & $0.3 \pm 0.2$ & $12.6 \pm 0.1$  
&202/139 &  $11 \pm 3$ & $0.5 \pm 0.2$ & -62\\
 \hline
\hline
6 
& $ 0.7 \pm 0.4 $ & $6.9 \pm 0.5 $ & $5.9 \pm 0.6$ 
& $160 \pm 30$ & $ 0.5 \pm 0.1$ & $18.1 \pm 0.2$ 
& 293/240 & $\leq 0.002 $ & 0.5 (f)& -0.3 \\
\hline
\end{tabular}
\tablefoot{
\tablefoottext{a}
{Plasma temperature of the Comptonized components.}
\tablefoottext{b}
{Plasma optical depth of the Comptonized components.}
\tablefoottext{c}
{Observed flux of the Comptonized components, in the quoted band.}
\tablefoottext{d}
{C-statistic for the unabsorbed model.}
\tablefoottext{e}
{Column density of the local neutral absorber.}
\tablefoottext{f}
{Covering fraction of the local neutral absorber.}
\tablefoottext{g}
{Decreasing of the C-statistic with respect to the unabsorbed model.}
  }
\label{var.tab}
\end{table*}


\subsection{Partial covering of the baseline model}
\label{spec.var}

Prompted by the results
of the low state fit,
we
checked if the addition of
a cold absorber could
improve also the
fit of the other datasets.
We outline
the results of
this exercise for
all the archival datasets
in Table \ref{var.tab}.
An additional 
partially-covering
absorbing component
with a similar covering factor
($\sim$ 50\%) but a lower
column density provides
a significant improvement
of the fit for Obs. 3--5. 
In contrast,
no absorbing component
is statistically required
in the fit of
Obs. 1 and
and 6, namely the two highest flux
states. For these datasets,
we set an upper limit to the
absorbing column density
by keeping the covering factor
fixed to $\sim$ 0.5.
In Fig. \ref{var_abs.fig} we plot
the parameters of the absorber
as a function of the source
flux. The absorbing column
density shows a negative
trend with the flux, while
the covering fraction
is consistent to be
constant.
The variable absorbing component
does not however account for
all the source variability.
As we show in Fig. \ref{var_lum.fig},
we still observe an intrinsic
variability, in both the
two Comptonized components,
after removing
the effect of the variable
absorption.\\
We tested also
a different scenario
where the absorber is constant
in opacity
and its variability is driven
only by a variable covering
fraction.
At this purpose, 
we attempted
to fit Obs. 4 and 5 keeping 
the absorber column density frozen 
to the value observed in the low-flux state 
and letting only the covering factor free to vary. 
In both cases we had to release 
also the parameters of the underlying continuum
to achieve an acceptable fit.
In detail,
for Obs. 4 we obtained a covering factor of $\sim$4\% 
but the model residuals are important between 2 and 5 keV. 
In Obs. 5 
the fit erases the
absorber pushing the covering factor 
to a much 
lower value ($\sim$0.03\%).
In both cases
the fit is statistically worse 
(C/d.o.f=216/152 and 152/139 respectively)
than what reported in Table \ref{var.tab}.
Thus, we rejected
this possibility,
and we concluded that
the absorber model outlined
in Table \ref{var.tab}
better fits the data.


\section{Discussion}
\label{disc}

\subsection{The X-ray spectrum of \th \source}

The long exposure \th \xmm \th \th
observation of \th \source \th \th 
that we presented in this paper
provided a high-quality X-ray
spectrum, suitable for testing
physically motivated models
against real data. Exploiting
the simultaneous coverage
in the optical/UV that
was provided in the
present case by the
OM and by HST-COS,
we successfully 
represented the broadband spectrum
of \th \source \th \th
using Comptonization.
The emerging physical
picture provided that
the optical-UV disk photons
($T \sim 56$ eV) are both Comptonized
by an optically thick ($\tau_{\rm wc} \sim 7$)
warm medium ($T_{\rm wc} \sim 0.7$ eV),
and by 
an optically thinner 
($\tau_{\rm hc} \sim 0.5$) 
and hotter ($T_{\rm hc} \sim 160$ keV) 
plasma to produce the entire
X-ray spectrum.\\
A similar
interpretation has been
recently proposed
for the broadband
simultaneous spectrum of
Mrk 509 (M11, P13),
and also for a sample of unobscured
type 1 AGN \citep{jin2012}.
A reasonable configuration
for these two media in the
inner region of AGN is
possible.
Two different Comptonizing 
coronae may 
be present.
The geometrically
compact hot corona may
be associated with
the inner part
of the accretion flow,
while the warm corona
may be a flat upper
layer of the accretion disk
(P13). Alternatively,
according to
a model proposed 
in \citet{don2012},
the warm Comptonization
may take place in the
accretion disk itself,
below a critical radius
after which the radiation
cannot thermalize anymore.
It is in principle also
possible that the seed
photons are provided to the
hot corona by the soft excess
component 
\citep[e.g., PKS 0558-504,][]{gli2013}. 
We note that,
provided a
slightly thicker warm Comptonized
component ($\tau \sim 11$), the broadband
spectrum of \th \source \th \th 
is consistent with
a ``nested-Comptonization''
scenario.
We suggest finally that
the \th \fek \th emission
line in \th \source \th \th
is produced
by reflection
in a cold thick torus. The long-term
flux stability of the line
that we note in Fig. \ref{var_fe.fig}
supports this interpretation.\\
\citet{tom2010} reported the detection of 
an ultra fast outflow (UFO) in \th \source. 
According to these authors, 
the signature of the UFO 
is a blushifted \ion{Fe}{xxvi}-\lia \th
absorption line located at a restframe energy 
of $\sim$7.23 keV, possibly accompanied by
a \ion{Fe}{xxv}
feature at $\sim$ 8.4 keV.
Despite the high signal to noise
none of these features is evident 
in the present spectrum. 
We estimated upper limits 
for the equivalent width ($EW$) 
of the main UFO absorption lines,
assuming the same outflow velocity 
given in \citet[][v$\sim$11100 \kms]{tom2010}.
The deepest UFO consistent with our
dataset is much shallower
than what previously reported
because we found $EW \la$ 12 eV
and $EW \la$ 9 eV for
the \ion{Fe}{xxvi}-\lia \th and
\ion{Fe}{xxv}-He$\alpha$ transition
respectively.

\begin{figure}[h]
\centering
\includegraphics[angle=90,width=0.5\textwidth,]{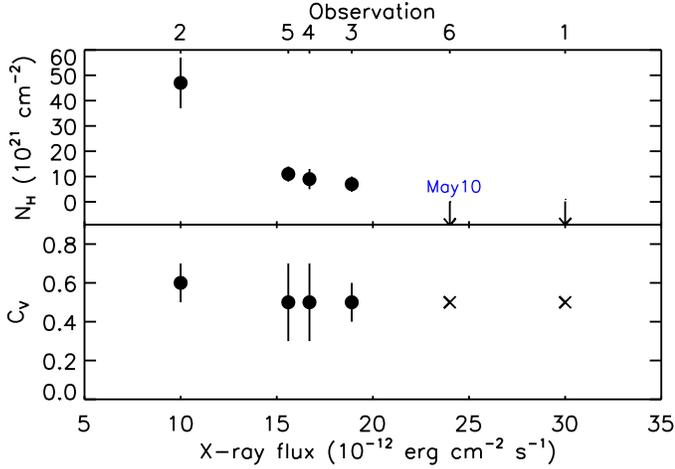}
\caption{
Variability of the local neutral absorber
in \th \source.
The gas column density (\textit{upper panel}) and
and covering factor (\textit{lower panel}),
with errors, are plotted against the observed source flux
in the 0.5--10.0 keV band. Arrows
represent upper limits for the column density.
Crosses represent fixed values for the covering factor.
The values for the
present dataset are labeled and the observation numbers for each data point
(see Table \ref{archive.tab})
are labeled as well on the upper axes.}
\label{var_abs.fig}
\end{figure}


\begin{figure}[h]
\centering
\includegraphics[angle=90,width=0.5\textwidth,]{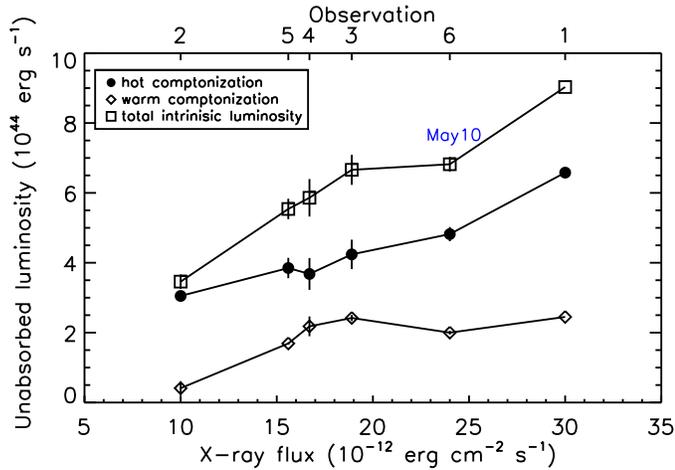}
\caption{
Variability of the intrinsic 0.5--10.0 keV luminosity in \th \source.
The unabsorbed luminosities of the hot and the warm Comptonized component,
along with the total intrinsic luminosities are plotted against
the observed source flux in the 0.5--10.0 keV band. The symbols
are outlined in the legend.
Error bars, when larger than the size of the plotted symbols,
are also shown. The values for the
present dataset are labeled and the observation numbers for each data point
(see Table \ref{archive.tab})
are labeled as well on the upper axes.}
\label{var_lum.fig}
\end{figure}


\subsection{The historical spectral variability of \th \source}

\th \source \th \th is well known for showing
a remarkable flux variability
in the X-rays
that is accompanied by
a dramatical flattening 
of the spectral slope
as the source flux
decreases.
Nonetheless
the optical/UV fluxes
that were observed
in diverse
X-ray flux states
are rather stable.
We showed in Sect. \ref{var}
that a Comptonization model
can explain
the historical spectral variability
of \th \source \th \th,
provided an intrinsic variability
of the two Comptonized components, 
and
a partially-covering,
cold absorption
variable in opacity.\\
The intrinsic variability
of the X-ray Comptonized continuum
is easy to justify.
Indeed, Monte Carlo computations
of X-ray spectra from
a disk-corona system
show that,  
even given an optical depth
and a electron temperature,
coronae may still be intrinsically
variable. Without requiring
variations in the accretion
rate that would be 
inconsistent with
the stability of the optical/UV
flux, a substantial variability
may be induced, for instance,
by geometrical variations
\citep[e.g.,][]{sob2004} or by
variations in the bulk velocity
of the coronal plasma
\citep[e.g., in a non static corona,][]{mal2001}.\\
However,
apart from the
smaller intrinsic
variation,
the variable
cold absorption 
causes the bulk 
of the observed
spectral variability.
In this respect,
\th \source \th \th
is not a unique case.
So far, cold
absorber variable
on a broad range of timescales
(few hours--few years)
have been
detected in a handful
of cases, 
\citep[e.g., NGC 1365, NGC 4388, and NGC 7582,][]{ris2005, elv2004, pic2007}
including
also some
optically unobscured,
standard type 1 objects
\citep[e.g., NGC 4151, 1H0557-385, Mrk 6][]{puc2007, lon2008, imm2003}
and narrow line Seyfert 1
\citep[e.g., SWIFT J2127.4+5654][]{san2013}.
In most of these cases,
discrete clouds of
cold gas, with densities
and sizes typical 
of the BLR clouds
may cause
the variable absorption.
The estimated location
of these clouds is within
the ``dust sublimation radius'',
nominally separating the BLR
and the obscuring torus.
This indicates that the distribution
of absorbing material in AGN
may be more complex
than the axisymmetric torus
prescribed by the standard Unified
Model \citep[see][and references therein]{bia2012}.\\ 
The variation of the cold absorber
in \th \source \th \th is driven
by a decreasing opacity,
which seems to show a trend
with the increasing flux
(Fig. \ref{var_abs.fig}, upper panel).
Indeed, in the two
highest flux spectra,
the absorbing column density
is consistent to have been at least
a factor $\sim$ 40 thinner than
in the lowest flux spectrum.
At the same time,
the absorber covering factor is
consistent to have remained
constant ($\sim$ 0.5).
A simple
explanation
for this behavior
is that in the
highest flux dataset,
the neutral
column density 
became ionized responding
to the enhanced X-ray
ionizing continuum.
An inspection of 
the archival
RGS spectra
\citep{pou2004a,pou2004b}
may provide
additional support to this
hypothesis. 
The broad emission
features from \ion{O}{vii}
and \ion{O}{viii} 
that we detected
in the RGS spectrum 
of the present dataset
(Paper I) are present
also in all the 
past RGS spectra.
We note however that
in the present dataset
the luminosity
of both these lines 
is lower than what observed,
for instance, in the
lowest flux state
(e.g. by a factor $\sim$2
and $\sim$5, respectively)
Moreover, we clearly detected
a broad line from
a more highly ionized
species (\ion{Ne}{ix}, see Paper I), 
that is undetected in all
the archival RGS spectra. 
This may
be a qualitative indication
of a more highly ionized
gas in the BLR during
the higher flux state.
The enhanced flux
may have increased
the ionization of the gas,
causing the enhancement of the
\ion{Ne}{ix} and the
decrease of the 
\ion{O}{vii}-\ion{O}{viii}.
The difference
in unabsorbed continuum
luminosities that we
observed for instance
between
the lowest flux state
of September 2002 and the
following observation of
March 2003
($\Delta L \sim 3 \times 10^{44}$ \ergs, see Fig. \ref{var_lum.fig}) 
may indeed ionize
a cloud of neutral
hydrogen with typical
BLR density in
the observed
timescale
($\Delta t \la $7 months).
Assuming that the absorber
is a cloud
of pure hydrogen that does
not change in volume as a consequence
of the ionization, then the fraction
of hydrogen that became 
completely ionized 
in March 2003:
\begin{equation}
\label{frach}
f_{\rm H}=\frac{{\it N}_{\rm H,Sep02}-{\it N}_{\rm H,Mar03}}{{\it N}_{\rm H,Sep02}} \sim 85\%.
\end{equation}
If the cloud is illuminated by $\Delta L$
the conservation of energy, 
assuming spherical symmetry
implies that:
\begin{equation}
\label{abs_en}
\frac{C_{\rm V} \Delta L \Delta t }{f \frac{4 \pi}{3} d^3} \sim U_{\rm H} f_{\rm H} n_{\rm H}.
\end{equation}
where:
$C_v$ is the absorber covering factor,
$d$ is the absorber distance,
$U_{\rm H}$=13.6 eV is the ionization threshold of hydrogen,
$n_{\rm H}$ is the absorber density, and
$f=10^{-2}$ is the volume filling factor
of the broad line region \citep{ost1989}.
Taking as un upper limit for the absorber
distance the dust sublimation
radius ($R_{\rm DUST}\sim$ 0.6 pc
\footnote{We estimated the dust sublimation radius
of \th \source \th \th using the ionizing luminosity given
in Paper I, and the formula of \citet{bar1987}, assuming
T=1500 K for the dust sublimation temperature 
and a=0.05 $\mu$m for the dust grains size.}), 
from Equation \ref{abs_en}
follows that:
\begin{equation}
\label{abs_den}
n_{\rm H} \ga 6 \times 10^8\rm \, cm^{-3}.
\end{equation}
This lower limit
is indeed well
consistent with
the typical range of densities
of the BLR clouds
\citep[$10^{8}-10^{12} \rm cm^{-3}$,][]{bal1995}.\\
We note also
that
this scenario
resulted in a statistically
better fit of
the data than
a model mimicking a
single cloud with constant
opacity crossing the line of sight
(see Sect. \ref{spec.var}).
Indeed,
the $\sim$months time interval
separating the \th \xmm \th
observations of \th \source \th \th
is inconsistent with the expected
duration of
an occultation event
due to a single
BLR cloud.
For instance,
in the case of NGC 1365 \citep{ris2007},
the occultation observed
in April 2006 lasted $\sim$4 days. 
A similar
eclipse, lasting
only  90 ks, has been observed
in SWIFT J2127.4+5654  \citep{san2013}.\\
In the scenario we are proposing
for \th \source,
when the continuum source 
is found in a low flux state,
the surrounding gas is on average less ionized. 
We suggest therefore that
in these conditions 
the number of neutral clouds
along the line of sight
may be larger,
making obscuration
events more probable.\\
We finally remark 
that in this framework,
a simple argument 
can explain the stability
of the optical/UV 
continuum in \th \source.
Indeed,
the optical/UV
continuum source
is 10 times larger in radius
than the X-ray source
\citep[see e.g.,][]{elv2012}.
Therefore,
a covering of the order of
$\sim50\%$ of the
X-ray source
implies a negligible covering
of the $\sim$ 0.5\%
for the optical/UV source.

\subsection{The geometry of AGN}

We can put significant constraints
on the geometry of \th \source \th \th
from the fitted parameters
of the disk blackbody
(Sect. \ref{opt.mod}).
Indeed,
the fitted disk
blackbody normalization
$A$ is linked to
the disk inclination
angle $i$. Analitically,
\begin{equation}
 A=R^2_{\rm in}\cos i,
\label{a}
\end{equation}
where
$R_{\rm in}$ is the inner radius 
of the disk.
The disk inner radius
may be set by the radius
of the innermost stable circular
orbit ($R_{\rm ISCO}$), which
is allowed in the space time metric
produced by the black hole mass
\citep[$M\sim3.8 \times 10^8$ \msun \th in this case,][]{one2005}
and by its spin. The two extreme
cases are a maximally rotating black
hole and a non-rotating Schwarzschild
black hole \citep[see e.g.,][]{bam2012}.
Therefore, according
to this general prescription,
$R_{\rm in}$ may vary only
in the range:
\begin{equation}
 R_{\rm in}\sim R_{\rm ISCO}=[1-6] \,R_{\rm g},
\label{risco}
\end{equation}
where
$R_{g}=2GM/c^2$
is the gravitational radius,
$G$ is the gravitational constant
and $c$ is the speed of light.
Combining equations \ref{a} and \ref{risco}
we obtain:
\begin{equation}
i=[70^{\circ}-89^{\circ}].
\end{equation}
To set a more robust lower limit
for the disk inclination angle,
we additionally considered in
the calculation the error
of the disk normalization
(Sect. \ref{opt.mod}) and of the
black hole mass \citep[$\sim$ 0.5 dex, see][and references therein]{one2005}.
With these tighter constraints:
\begin{equation}
i \geq 54^{\circ}.
\end{equation}
These inclination 
values are well consistent
with the intermediate
spectral classification
of \th \source \th \th as
Seyfert 1.5. It is
therefore possible
that our viewing direction
towards \th \source \th \th
is grazing the
so called ``obscuring torus''
prescribed by the
standard Unified 
Model \citep{ant1993}.
In this framework,
the inner part of the torus
may be responsible for the
X-ray obscuration.\\
The structure of the obscuring
medium in AGN may be more complex
than the classical
donut torus paradigm.
Indeed, this model
faces difficulties
in explaining several theoretical
and observational issues,
including for instance the wide range of
X-ray obscuring column densities
\citep[see][and references therein]{elv2012}.
A clumpy torus \citep{nen2008} 
may alleviate part of these problems.
In the latter case,
when the numerical
density of clouds along
the line of sight is low,
even a source viewed from a 
high inclination
angle may appear like
a Seyfert 1
\citep[see also][]{eli2012}.
Moreover,
in this model,
the BLR and the
torus itself
are
part of the same medium, 
decreasing in ionization
as the distance from the
central source increases.
Indeed,
it has been proposed
\citep[see][]{nen2008}
that
the clumpy torus 
extends inward beyond the dust sublimation 
point.
The innermost
torus clouds,
being more exposed to
the ionizing radiation,
are probably dust free
and may dominate the 
X-ray obscuration.
Given the high
viewing angle derived
above,
\th \source \th \th
may possibly
fit in this framework.

\subsection{A comparison with other models}

The model 
that we presented in this paper
explains both 
the optical/UV/X-ray broadband spectrum,
and the historical variability of
\th \source in a reasonable
geometrical configuration. 
However,
the hard X-ray energy range
above 10 keV is not covered
by the present spectral analysis.
In that range, \th \source \th \th
displays a ``hard-excess''
\citep{tur2009, pal2013, wal2010}
over a simple power law model, 
that shows some evidence
of variability 
\citep[a factor $\sim$ 2, see][]{tur2009}.
The extrapolation to harder energies of
our broadband model predicts a flux of
$\sim2.9 \times 10^{-11}$ \ergsc
in the 10--50 keV band. 
This flux is
higher than the
70 months-averaged flux observed
with BAT in the same band
\citep[$\sim2.2 \times 10^{-11}$ \ergsc,][]{bau2013},
but well consistent
with the latest
$Suzaku$
measurement
taken only 5 months
before our \th \xmm \th \th
observations 
\citep[$\sim2.7 \times 10^{-11}$ \ergsc,][]{pal2013}.
Because of this relatively short time interval 
between the $Suzaku$ and \th \xmm \th \th observation,
it is indeed likely that
$Suzaku$ caught
the source in the
same flux condition
of our observation
(see also Fig. \ref{fluxes.fig}).
In the context
of the $Suzaku$ data analysis
\citep{pal2013},
the authors attempted also
to fit
a Comptonization
model to the May 2010 EPIC-pn
spectra.
They obtained however
a poor result
mainly because
of a prominent excess
in the residuals
near $\sim$ 0.5 keV.
In the present analysis,
thanks to the higher resolution
provided by the RGS,
we could easily identify that
feature
as due to the \ion{O}{vii} line complex.
Apart from this discrepancy, the
parameters they obtained for
the warm Comptonized component
reasonably agree with our result.\\
On a different occasion
(July 2007), $Suzaku$ caught
the source in a bright
state, similar to what
was observed by \th \xmm \th \th
in December 2000 (Obs.1).
The hard X-ray curvature 
that was observed in that case
($F_{15-50 \rm \, keV} \sim 2.6 \times 10^{-11}$ \ergs)
can be fitted using
a Compton-thick,
highly ionized absorption,
covering $\sim$66\% of the line of sight.
To check if our model could
explain also this historical
hard X-ray maximum,
we made the exercise
of comparing 
the hard X-ray flux extrapolated
by the fit of Obs. 1 with the one
observed by $Suzaku$.
We noted a small
disagreement, with
the flux predicted
by our model being 
by a factor $\sim$ 1.6 
lower than the observed one. 
Thus, we cannot definitively rule out that
a partially-covering ionized absorber
was present in the high flux state
of July 2007. In the
framework proposed in this
paper, it may in principle 
be the ionized counterpart of the cold
absorber present in the low
flux state.\\
Beside our interpretation,
also the light bending model 
may explain
the variable X-ray spectrum
of this source and fits both
the \th \xmm \th \th and the $Suzaku$
datasets \citep[F05, ][this paper]{wal2010,pal2013}.
We note however that
even in this disk-reflection
framework,
a variable absorption
is required to fit the data.
Indeed, a cold absorber 
showing the same trend
noticed here
(with a column density
dropping from \nh$\simeq 10^{21}$ \colc to 0) 
is present in the F05
model.
Additionally,
an \ion{O}{vii}
edge with a variable depth, 
mimicking an ionized warm absorber
is also included.
The latter
is somewhat at odds with
what we reported in Paper I
because the
warm absorber in \th \source \th \th
is too lowly ionized to produce
any strong \ion{O}{vii} absorption features.
Moreover, a short timescale variation of the
ionized absorption edges, as required in F05,
is difficult to reconcile
with the galactic scale location
(see Paper I) of the warm absorber in this source.
In our analysis, solar abundances are
adequate to fit the data.
In the light bending
model, the metal abundance
in the disk is a free parameter,
and it has been reported to
vary from supersolar in
September 2002 ($\sim3.8$, F05)
to 
undersolar in the January 2010
\citep[$\sim$0.5,][]{pal2013}.
This is another issue that
is difficult to explain.
Finally, we note that a
physical interpretation of the
source variability observed with $Suzaku$
is not possible in the context of the
light bending model alone, and
additional variability
in the disk-corona geometry,
possibly caused by a variability
in the accretion rate
(that would be however inconsistent
with the stability of the optical/UV
flux noticed here)
has to be invoked \citep{pal2013}.\\
In conclusion,
the Comptonization model proposed
here for the present and historical
broadband spectrum of \th \source \th \th
does not rule out other
possible interpretations. It has
however the advantage of explaining
all the observational evidences
collected in the last ten years
over the broadest range
of wavelength available,
without requiring any
special \textit {ad hoc} assumption.
A future observation 
in the entire X-ray band
with XMM and NuSTAR would be
important
to solve
the long-standing ambiguity 
in the interpretation
of the spectral variability 
in this peculiar Seyfert galaxy.
%
%
%
\section{Summary and conclusions}
\label{conc}

We modeled the broadband
optical (XMM-OM), UV (HST-COS, FUSE)
and X-ray (EPIC-pn) simultaneous
spectrum of \th \source \th \th
taken in May 2010 using Comptonization.
The X-ray continuum may be produced
by a warm ($T_{\rm wc} \sim 0.7 $ keV, $\tau_{\rm wc} \sim 7 $) 
and a hot Comptonizing medium
($T_{\rm hc} \sim 160 $ keV, $\tau \sim 0.5$)
both fed by the same optical/UV
disk photons ($T_{\rm dbb} \sim 56$ eV).
The hot medium may be
a geometrically compact
corona located in the
innermost region of the
disk.
The warm medium 
may be an upper layer
of the accretion disk.
Reflection
from cold distant matter
is a possible origin 
for the \th \fek \th emission
line. Despite the
long exposure time of our
dataset we do not
find evidence of the
ultra fast outflow features
that have been reported in the
past for this source. \\ \\
Providing a partially
covering ($\sim$50\%) cold absorber
with a variable opacity
(\nh $\sim [10^{19}- 10^{22}]$ \colc)
and a small 
variability intrinsic to the source,
this model can reproduce
also the historical spectral
variability of \th \source.
The opacity of the absorber
increases as the continuum
flux decreases.
We argue that
the absorber may have the typical
density of the BLR clouds
and that it,
getting ionized in response
to the enhanced X-ray continuum,
becomes optically thinner in the
higher flux states.\\ \\
Relativistic light-bending 
remains
an alternative explanation
for the spectral variability
in this source. We note
however that
in this scenario,
a variable elemental
abundance and a variable
absorption
are required.
The latter is difficult
to reconcile with the UV/X-rays
absorber that we have determined
to be located at a $\sim$ kpc
scale.\\ \\
Finally, we suggest that
\th \source \th \th may be viewed
from a high inclination
angle, marginally intercepting 
a possibly clumpy obscuring torus.
In this geometry, the X-ray
obscuration may be associated
with the innermost dust free
region of the obscuring torus.\\ \\
The present spectral analysis
in the optical/UV/X-ray
represents a substantial step
forward
in the comprehension
of this intriguing Seyfert
galaxy. However,
further investigations
(e.g. with NuSTAR)
are needed to 
understand the true nature
of the spectral variability
of this source.
%

\begin{acknowledgements}
This work is based on observations with XMM-Newton, 
an ESA science mission with instruments and contributions 
directly funded by ESA Member States and the USA (NASA). 
SRON is supported financially by NWO,
the Netherlands Organization for Scientific Research.

\end{acknowledgements}


\end{document}